\documentclass[runningheads]{llncs}
\usepackage[T1]{fontenc}
\usepackage{amsfonts,amssymb}
\usepackage{multirow}
\usepackage{graphicx}
\usepackage{mathrsfs}
\usepackage{caption}
\DeclareCaptionLabelSeparator{dot}{. }
\usepackage{subcaption}
\usepackage{xcolor}
\usepackage{hyperref}
\usepackage[misc]{ifsym}
\usepackage{booktabs}
\hypersetup{
  colorlinks=true,
  linkcolor=blue, 
  citecolor=blue, 
  urlcolor=blue, 
  linkbordercolor=white, 
  citebordercolor=white
}
\begin{document}
\title{NexToU: Efficient Topology-Aware U-Net for Medical Image Segmentation}
\author{Pengcheng Shi\inst{1} \and Xutao Guo\inst{1,2} \and Yanwu Yang\inst{1,2} \and \\Chenfei Ye\inst{4} \and Ting Ma\inst{1,2,3,4}$^{(\textrm{\Letter})}$}
\authorrunning{P. Shi et al.}
\institute{Electronic \& Information Engineering School, Harbin Institute of Technology (Shenzhen), Shenzhen, China \and
Peng Cheng Laboratory, Shenzhen, China \and
Guangdong Provincial Key Laboratory of Aerospace Communication and Networking Technology, Harbin Institute of Technology (Shenzhen), Shenzhen, China \and International Research Institute for Artificial Intelligence, Harbin Institute of Technology (Shenzhen), Shenzhen, China}

%\institute{Electronic \& Informatin Engineering School, Harbin Institute of Technology (Shenzhen), Shenzhen, China\\
%\email{tma@hit.edu.cn} \and Peng Cheng Laboratory, Shenzhen, China \and Guangdong Provincial Key Laboratory of Aerospace Communication and Networking Technology, Harbin Institute of Technology (Shenzhen), Shenzhen, China \and International Research Institute for Artifcial Intelligence, Harbin Institute of Technology (Shenzhen), Shenzhen, China}
%
\maketitle              % typeset the header of the contribution
\begin{abstract}
Convolutional neural networks (CNN) and Transformer variants have emerged as the leading medical image segmentation backbones. Nonetheless, due to their limitations in either preserving global image context or efficiently processing irregular shapes in visual objects, these backbones struggle to effectively integrate information from diverse anatomical regions and reduce inter-individual variability, particularly for the vasculature. Motivated by the successful breakthroughs of graph neural networks (GNN) in capturing topological properties and non-Euclidean relationships across various fields, we propose NexToU, a novel hybrid architecture for medical image segmentation. NexToU comprises improved Pool GNN and Swin GNN modules from Vision GNN (ViG) for learning both global and local topological representations while minimizing computational costs. To address the containment and exclusion relationships among various anatomical structures, we reformulate the topological interaction (TI) module based on the nature of binary trees, rapidly encoding the topological constraints into NexToU. Extensive experiments conducted on three datasets (including distinct imaging dimensions, disease types, and imaging modalities) demonstrate that our method consistently outperforms other state-of-the-art (SOTA) architectures. All the code is publicly available at \url{https://github.com/PengchengShi1220/NexToU}.

\end{abstract}
\section{Introduction}
Accurate medical image segmentation is essential for clinical practice, enabling quick identification and analysis of regions of interest (ROI) in images. In clinical applications, it plays a critical role by providing valuable information about a patient's anatomy and physiology.
\begin{figure}[htbp]
\centering
\includegraphics[width=\textwidth]{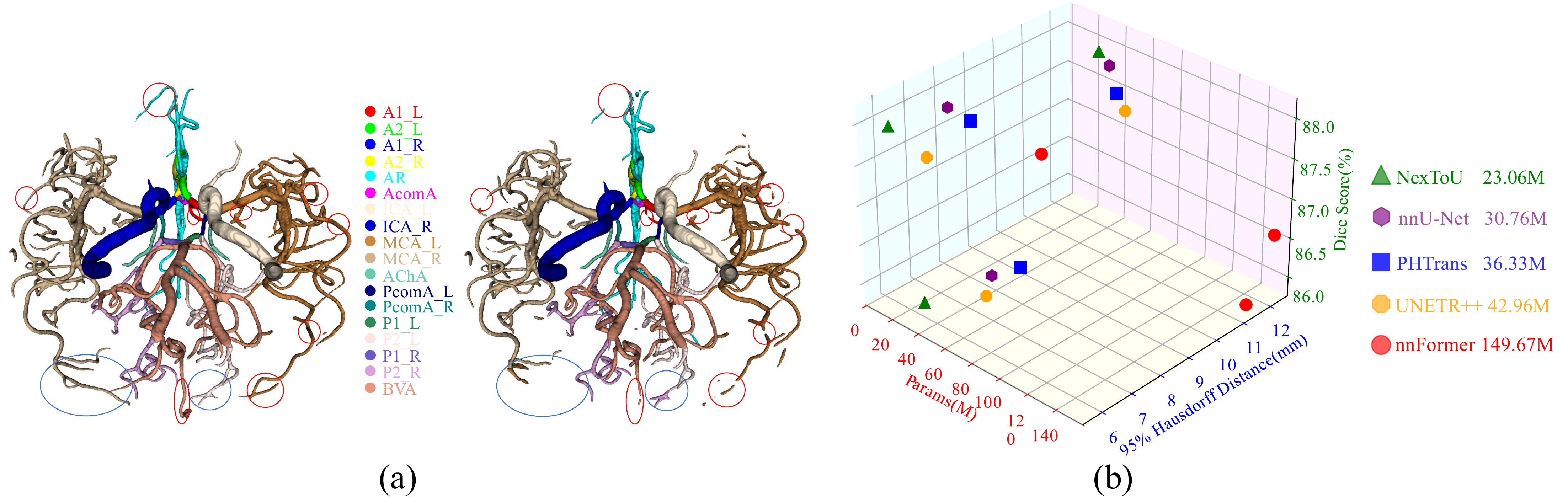}
\caption{(a) The left illustration shows the ground truth (GT) of intracranial arteries (ICA), while the right illustration displays the predicted results from PHTrans \cite{liu2022phtrans}. (b) The performance comparison on the BTCV dataset in three-dimensional projection. (Color figure online)} \label{fig1}
\end{figure}
Especially for the brain vasculature, achieving precise segmentation of individual vasculature is an important initial step towards the standardized and non-invasive analysis of cerebral arteries \cite{dumais2022eicab}, which contributes to identifying Alzheimer's disease (AD) or stroke in advance \cite{paetzold2021whole}. \\
\indent{Thanks to the advancement of deep learning in the field of vision, convolutional neural networks (CNN) and Transformer variants have been the leading backbones in medical image segmentation. Nevertheless, CNN variants \cite{ronneberger2015u,zhou2018unet++} and transformer variants \cite{chen2021transunet,cao2023swin} still have several drawbacks when segmenting irregular objects. Although CNN variants possess inherent inductive biases that enable them to generalize well on small datasets \cite{han2022survey}, they cannot efficiently capture global context information. Transformer variants have computational redundancy when relying purely on attention mechanisms to learn global dependencies. Despite recent studies that proposed hybrid architectures combining the advantages of CNN and Transformer \cite{zhou2021nnformer,liu2022phtrans}, the challenge of segmenting complex anatomical structures while maintaining both efficiency and accuracy persists. Fig.~\ref{fig1} (a) shows a case of ICA multi-class segmentation results compared with the GT. It depicts that certain connections between small vessels are interrupted (red circle) and spatial confusion in labeling violates the anatomy of the ICA (blue circle).}\\ 
\indent{Graph neural networks (GNN) are specifically designed for tasks that involve naturally constructed graphs, such as social networks, biological networks, and citation networks. Meanwhile, recent advances have shown the potential of GNN for medical image analysis \cite{paetzold2021whole,chen2020automated,bongratz2022vox2cortex,yang2020cpr} and GNNs have unique advantages in tubular structure segmentation \cite{shin2019deep}. Vox2Cortex \cite{bongratz2022vox2cortex} and CPR-GCN \cite{yang2020cpr} demonstrated promising results in utilizing the structured and networked properties of medical images for cortical surface reconstruction and automated anatomical labeling of the coronary artery, respectively. The hybrid CNN-GNN architecture proposed by Shin et al. \cite{shin2019deep} improves vessel segmentation accuracy by leveraging both local appearances and global vessel structures. Furthermore, the ability of Vision GNN (ViG) to process image data directly \cite{han2022vision} has facilitated its achievement of state-of-the-art (SOTA) performance in various vision tasks, we argue that hybrid architectures using ViG variants can efficiently correlate information from various anatomical regions and extract tubular structures.\\
\indent{In this paper, we introduce a novel hybrid architecture for complex anatomical structures segmentation in medical images, dubbed as \textbf{Ne}twork with \textbf{x} \textbf{To}pological modules in a \textbf{U}-shape: \textbf{NexToU}. NexToU connects nodes across different anatomical regions from a graph perspective. To improve the topological representations, the approach involves employing the Pool GNN module and Swin GNN module, based on the Swin Transformer \cite{liu2021swin}. The Pool GNN module identifies key nodes in the global network to extract crucial topological information, while the Swin GNN module captures local information through shifted window partition and reverse operations, making it effective for recognizing irregularly shaped vasculature. We also incorporate a binary topological interaction (BTI) module, derived from the topological interaction (TI) module \cite{gupta2022learning}, into the loss function calculation to differentiate anatomical structures based on their topological relationships at adjacent locations. Our contributions are summarized as 1) We introduce NexToU, the first convolutional ViG-based architecture for medical image segmentation, utilizing two Efficient ViG (EViG) modules (Pool GNN and Swin GNN) to aggregate global and local topological representations in the latent space. 2) BTI is presented to decrease the number of convolution operations in the original TI module from $c\cdot(c-1)$ to $2\times(c-1)$ for the c-class medical image segmentation task. 3) Comprehensive experiments validate that our approach outperforms other competing methods on three distinct medical image segmentation tasks. Fig.~\ref{fig1} (b) shows the performance comparison on the BTCV dataset and NexToU achieves a higher Dice Score Coefficient (DSC) and smaller Hausdorff Distance (HD) with fewer parameters.}
\section{NexToU}
The proposed NexToU architecture follows a hierarchical U-shaped encoder-decoder structure \cite{ronneberger2015u}  including pure convolution modules and $x$ topological ones. An overview of the NexToU architecture is illustrated in Fig.~\ref{fig2}.
\begin{figure}[htb]
\centering
\includegraphics[width=\textwidth]{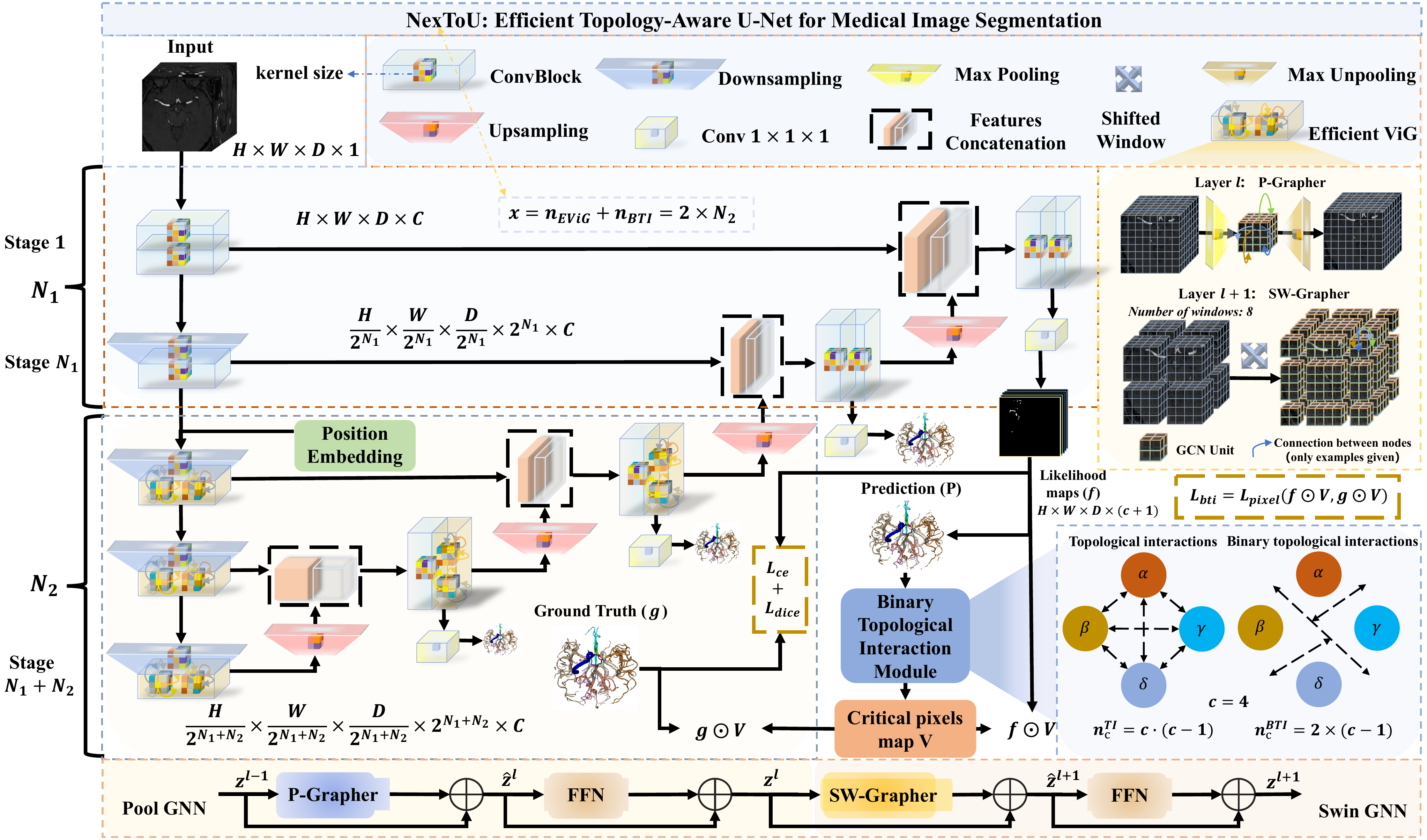}
\caption{Overview of the proposed NexToU architecture. It mainly comprises two components: 1) Two EViG modules learning both global and local topological representations. 2) BTI module rapidly encodes topological constraints for multi-class medical image segmentation. (Color figure online)} \label{fig2}
\end{figure}
\subsection{Efficient Vision GNN}
Two novel EViG modules, namely Pool GNN and Swin GNN, are specifically designed to extract topological features. Pool GNN comprises of two parts: P-Grapher and FFN modules, and similarly, Swin GNN consists of SW-Grapher and FFN modules. Subsequently, we select the volumetric (3D) segmentation task to present these two EViG modules. For volumetric segmentation, assuming that the 3D input (volume) $\textbf{X}\in\mathbb{R}^{H\times W\times D\times C}$ with resolution $(H,W,D)$, C base channels and B batch size, we first obtain feature maps $\textbf{F}\in\mathbb{R}^{\frac{H}{2^h}\times\frac{W}{2^w}\times\frac{D}{2^d}\times2^{N_1}\times C}$ utilizing $N_1$ pure convolution modules, where $h=\sum_{i=1}^{N_1}p_{i}^{h}$, $w=\sum_{i=1}^{N_1}p_{i}^{w}$, $d=\sum_{i=1}^{N_1}p_{i}^{d}$, $p_{i}^{h}$/$p_{i}^{w}$/$p_{i}^{d}$ is a variable that signifies whether the $h$/$w$/$d$ variable performs a pooling operation at the $i^{\rm th}$ convolution module, with 0 and 1 representing no pooling and pooling, respectively. Also, $(\frac{H}{2^h},\frac{W}{2^w},\frac{D}{2^d})$ is the resolution of each patch and $N=\frac{H}{2^h}\times\frac{W}{2^w}\times\frac{D}{2^d}$ denotes the length of the sequence. We further augment $\textbf{F}$ by integrating a learnable position embedding after $N_1$ pure convolution modules. By transforming each patch into a feature vector $\textbf{X}_i\in\mathbb{R}^{D_{\rm v}}$, we have $X=\lbrack \textbf{X}_1,\textbf{X}_2,\cdots,\textbf{X}_N\rbrack$ where $D_{\rm v}=N\times2^{N_1}\times C$, $i=1,2,\cdots,N$ and $2^{N_1}\times C$ is the feature dimension. To efficiently perform global and local topological representations, we incorporate two modules, P-Grapher and SW-Grapher, after the convolutional layer.\\
{\bfseries P-Grapher module.} At layer $l$, we reduce the computational redundancy of the original Grapher module in global topological representations by employing a novel P-Grapher module. This involves using max-pooling and max-unpooling operations before and after the Grapher module. After applying the maximum pooling operation, the resolution of each patch is $(\frac{H}{2^{h+1}},\frac{W}{2^{w+1}},\frac{D}{2^{d+1}})$, and the length of the sequence $N_{\rm p}=\frac{H}{2^{h+1}}\times\frac{W}{2^{w+1}}\times\frac{D}{2^{d+1}}$. By transforming each patch into a feature vector $\textbf{X}_i\in\mathbb{R}^{D_{\rm pv}}$, we have $X_{\rm p}=\lbrack \textbf{X}_1,\textbf{X}_2,\cdots,\textbf{X}_{N_{\rm p}}\rbrack$ where $D_{\rm pv}={N}_{\rm p}\times2^{N_1}\times C$. In particular, the P-Grapher module refrains from performing the pooling operation during the final two steps to mitigate the loss of deep semantic information. This is identical to the original Grapher module when the patch resolution remains at $(\frac{H}{2^h},\frac{W}{2^w},\frac{D}{2^d})$ and $N_{\rm p}=N,D_{\rm pv}=D_{\rm v},X_{\rm p}=X$. Then we view these features as a set of unordered nodes which are denoted as $\mathcal{V_{\rm p}}=\lbrace v_1,v_2,\cdots,v_{N_{\rm p}}\rbrace$. For each node $v_i$, we follow \cite{li2019deepgcns} to find its $K$ nearest neighbors $\mathcal{N}_{\rm p}(v_i)$ and add an edge $e_{ji}$ directed from $v_j$ to $v_i$ for all $v_j\in \mathcal{N}_{\rm p}(v_i)$. Subsequent to this, we acquire a graph $\mathcal{G_{\rm p}}=(\mathcal{V_{\rm p}},\mathcal{E_{\rm p}})$ where $\mathcal{E_{\rm p}}$ denote all the edges. We refer to the graph construction process as $\mathcal{G_{\rm p}}=G(X_{\rm p})$ in the following. By treating the image as graph data, we investigate the use of GNN to extract its representation, where the dynamic maximum graph convolutional processing is based on ViG \cite{han2022vision}. Multi-head update operation within dynamic graph convolution (DGC) enables diverse feature processing by updating information across multiple subspaces. \\
{\bfseries SW-Grapher module.} We adapt the Swin Transformer \cite{liu2021swin} by introducing the shifted window partition and window shifted reverse operations at the start and end of the S module, respectively. The shifted window partition divides the input volumes into a sequence of non-overlapping windows with a specified size, while the shifted window reverse performs the inverse operation. This enables efficient modeling of token interactions at layer $l+1$ through local graph convolution within each region. Specifically, we use a window of size ${N_{\rm m}}=M\times M\times M$ to evenly divide a 3D token into ${N_{\rm w}}=\lceil\frac{H}{2^h\times M}\rceil\times\lceil\frac{W}{2^w\times M}\rceil\times\lceil\frac{D}{2^d\times M}\rceil$ windows, producing feature maps $\textbf{F}_{\rm w}\in\mathbb{R}^{N_{\rm m}\times 2^{N_1}\times C}$ and batch size $B_{\rm w}=B\times N_{\rm w}$. After transforming each window into a feature vector $\textbf{X}_i\in\mathbb{R}^{D_{\rm wv}}$, we have $X_{\rm w}=\lbrack \textbf{X}_1,\textbf{X}_2,\cdots,\textbf{X}_{N_{\rm m}}\rbrack$ where $D_{\rm wv}=N_{\rm m}\times 2^{N_1}\times C$, $i=1,2,\cdots,N_m$ and $2^{N_1}\times C$ is the feature dimension. Subsequently, we consider these features as a set of unordered nodes, denoted as $\mathcal{V}_{\rm w}=\lbrace v_1,v_2,\cdots,v_{N_{\rm m}}\rbrace$, and the process of graph construction $\mathcal{G_{\rm w}}=G(X_{\rm w})$ is following the same steps as the P-Grapher module. Besides, we shift the partitioned windows by $(\lceil\frac{M}{2}\rceil,\lceil\frac{M}{2}\rceil,\lceil\frac{M}{2}\rceil)$ voxels. The shifted windowing mechanism is illustrated in Fig.~\ref{fig2}. The outputs of encoder blocks in layers $l$ and $l+1$ are computed as:
\begin{equation}\label{eq:1}\hat{z}^l={\rm P\rm{-}Grapher}(z^{l-1})+z^{l-1},\quad z^l={\rm FFN}(\hat{z}^l)+\hat{z}^l\end{equation}\begin{equation}\label{eq:2}\hat{z}^{l+1}={\rm SW\rm{-}Grapher}(z^l)+z^l,\quad z^{l+1}={\rm FFN}(\hat{z}^{l+1})+\hat{z}^{l+1}\end{equation} where P-Grapher and SW-Grapher denote pooling and shifted window partitioning Grapher modules, respectively, $\hat{z}^l$ and $\hat{z}^{l+1}$ are the outputs of P-Grapher and SW-Grapher; FFN denotes Feed Forward Network (see Fig.~\ref{fig2}).}
\subsection{Binary Topological Interaction Module}
As mentioned in \cite{gupta2022learning}, prior approaches overlook the topological interactions among labels and focus solely on pixel accuracy. Although \cite{gupta2022learning} shows experimental evidence of the efficacy of the topological interaction (TI) module, its efficiency declines with an increasing number of classes. We introduce the BTI module in this study to reduce the number of convolution operations in the original TI module for global topological constraints in medical image segmentation tasks. To obtain a topological critical pixel map $V$ for two different classes (or a merged new class), we apply $n_{c}^{\rm TI}$ or $n_{c}^{\rm BTI}$ convolution operations using the main workflow of the TI module with $d$ divisions, where each division involves convolution operations between two classes, resulting in a total of $2d$ convolution operations.\\
\indent{Based on the nature of binary trees, it can be stated that for any non-empty binary tree with $n_0$ leaf nodes and $n_2$ nodes of degree 2. During the division process of the BTI module shown in Fig.~\ref{fig2}, the number of leaf nodes is equal to the number of classes $c$, while the number of nodes of degree 2 is equal to $d$. Thus, we can obtain Eq. \ref{eq:4} through derivation. According to the anatomical structure of each dataset, we have designed the BTI module to minimize computational complexity. The supplemental materials detail the binary tree structure for each category in each dataset and provide a comparison of the TI module and BTI module single-epoch runtimes under identical conditions across different datasets. \begin{equation}\label{eq:3}n_{c}^{\rm TI}=2\cdot C^2_c=c\cdot(c-1)\end{equation}\begin{equation}\label{eq:4}n_2=n_0-1,\quad d=c-1,\quad n_{c}^{\rm BTI}=2\times(c-1)\end{equation}
\subsection{Loss Function}
We improve performance using $L_{\rm pixel}$ as the pixel-wise loss function, following \cite{gupta2022learning}. To ensure topologically correct segmentation, we introduce the binary topological interaction loss, $L_{\rm bti}$, which utilizes the critical pixels map $V$, likelihood maps $f$, and ground truth $g$ (Eq. \ref{eq:5}). The final loss function, $L_{\rm total}$ (Eq. \ref{eq:6}), combines the cross-entropy and dice loss with adjustable weights $\lambda_{\rm dice}$ and $\lambda_{\rm bti}$. For 2D tasks, $\lambda_{\rm bti}$ is 1e-4, while for 3D tasks, it’s 1e-6. \begin{equation}\label{eq:5}L_{\rm bti} = L_{\rm pixel}(f \odot V, g \odot V)\end{equation} \begin{equation}\label{eq:6}L_{\rm total} = L_{\rm ce} + \lambda_{\rm dice}L_{\rm dice} + \lambda_{\rm bti}L_{\rm bti}\end{equation}
\section{Experiments and Results}
\subsection{Dataset and Implementation Details}
We thoroughly evaluated the effectiveness of our proposed NexToU architecture on three segmentation tasks. For the first task, we performed ICA multi-class segmentation on brain MRA datasets from OASIS-3 \cite{lamontagne2019oasis}, categorizing the segments into 18 distinct classes according to \cite{chen2020automated}. Each image was annotated by a clinical expert and double-blind reviewed by two specialists. The supplementary material provides a schematic illustration of the ICA. We utilized a 3D input with a patch size of 48×224×224, and randomly split 80 images with a ratio of 7:1:2 for training, validation, and testing, while removing mirroring from the data augmentation \cite{wasserthal2022totalsegmentator}. The second task involved abdominal multi-organ segmentation using the Beyond the Cranial Vault (BTCV) abdomen challenge dataset \cite{landman2015miccai} with a patch size of 48×192×192. The evaluation metrics used were average DSC and average HD, and the method was evaluated on eight abdominal organs. Following \cite{zhou2021nnformer}, we split the dataset into 18 training samples and 12 testing samples. For the third task, we conducted segmentation of retinal arteries and veins using the Retinal Arteries and Veins in Infrared Reflectance (RAVIR) imaging challenge dataset \cite{hatamizadeh2022ravir,hatamizadeh2020artificial}. We selected 18 images for training and 5 images for testing from RAVIR and resized all images to 384×384.\\
\indent{NexToU was implemented using PyTorch 1.11 and trained on a single NVIDIA-V100 GPU. To maintain consistency in data augmentation and post-processing, 
\begin{figure}[htb]
\centering
\includegraphics[width=\textwidth]{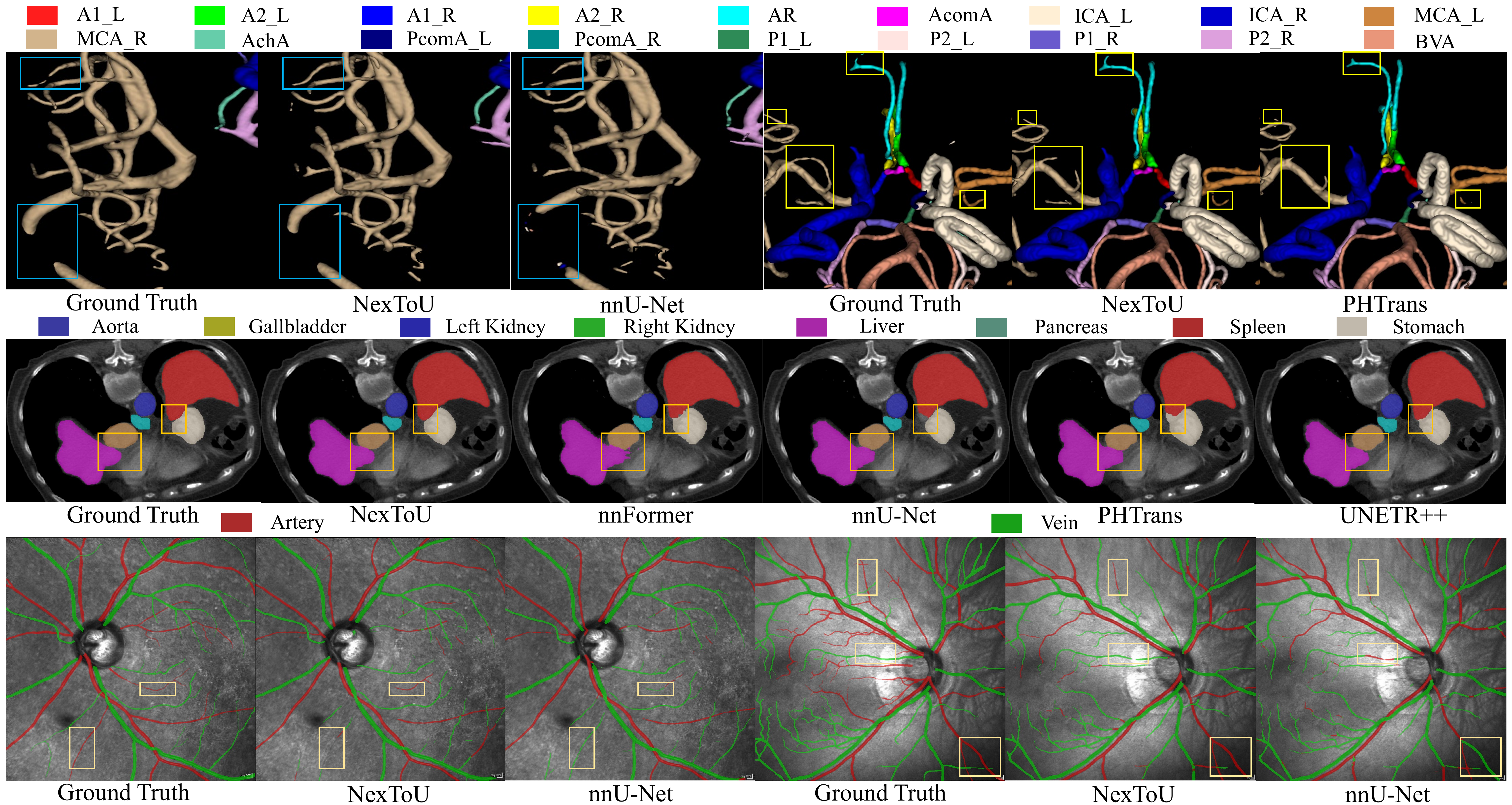}
\caption{Qualitative visualizations of the proposed NexToU and other methods.} \label{fig3}
\end{figure}
we base our NexToU architecture on the nnU-Net framework \cite{isensee2021nnu} which can automatically configure itself for any new medical image segmentation task. For ICA and abdominal multi-class segmentation, a batch size of 2 was used, while for RAVIR dataset, a batch size of 6 was selected. We set C=24 as the base channel. For $N_2$=3, the number of K of EViG used in different encoder stages was set to [4, 8, 16]. Additionally, we used h=6 for 3D and h=4 for 2D as the default number of heads in the multi-head update operation.  Furthermore, we utilized DGC of Swin Grapher with batch normalization and instance normalization for the Pool Grapher's DGC.}
\subsection{Comparisons with the State-of-the-Art Methods}
Extensive experiments were conducted on three datasets using various modalities in both 2D and 3D settings, which demonstrate that NexToU surpasses SOTA methods in medical image segmentation (Fig.~\ref{fig3}). On the BTCV dataset, NexToU achieved an average DSC of 87.84\% and an average HD of 6.33 mm. It also achieved the highest DSC of 74.19\% on the ICA dataset and showed promising results on the RAVIR dataset as presented in Table~\ref{tab1}, Table~\ref{tab2}, and Table~\ref{tab3}, respectively. NexToU outperforms alternative models in inter-class boundary segmentation and parameter efficiency (Fig.~\ref{fig1}). Compared to other methods generates superior segmentation results with fewer misclassifications (e.g., spleen, liver, stomach, and middle cerebral artery). Overall, NexToU achieves superior performance in medical image segmentation, particularly for tubular structures, by learning both global and local topological representations. Furthermore, it holds great promise for advancing medical image analysis. %\noindent
\begin{table}
\centering
\caption{Segmentation results for the ICA dataset (averaged over arteries with left and right hemispheres. AA=AcomA, PA=PcomA, average DSC \%).}\label{tab1}
\begin{tabular}{c|c|ccccccccccc}
\hline
Methods & DSC$\uparrow$ & A1 & A2 & AR & AA & ICA & MCA & AchA & PA & P1 & P2 & BVA \\
\hline
nnU-Net \cite{isensee2021nnu}& 73.59 & 82.33 & 73.07 & 77.05 & 52.06 & 92.28 & 79.67 & 56.38 & 66.68 & 73.52 & 72.04 & 84.40\\
nnFormer\cite{zhou2021nnformer}& 72.66 & 78.85 & 70.85 & 80.41 & 48.49 & 93.18 & 82.53 & 54.18 & 64.88 & 74.89 & 66.15 & 84.82\\
PHTrans \cite{liu2022phtrans}& 73.16 & 80.34 & 70.31 & 79.10 & 47.56 & 92.00 & 82.83 & 56.56 & 65.34 & 74.79 & 70.37 & 85.53\\
\hline
NexToU & \textbf{74.19} & 79.53 & 72.28 & 78.12 & 54.20 & 91.30 & 83.99 & 59.50 & 63.97 & 75.82 & 72.78 & 84.57\\
\hline
\end{tabular}
\end{table}

\begin{table}
\centering
\caption{Results of segmentation on the BTCV dataset.}\label{tab2}
\begin{tabular}{c|cc|cccccccc}
\hline
Methods & DSC$\uparrow$ & HD$\downarrow$ & Aot & Gal & Kid (L) & Kid (R) & Liv & Pan & Spl & Sto\\
\hline
nnU-Net \cite{isensee2021nnu}& 87.71 & 8.28 & 93.02 & 73.21 & 87.86 & 88.82 & \textbf{97.27} & 83.47 & 92.35 & 85.68\\
TransUNet \cite{chen2021transunet}& 77.48 & 31.69 & 87.23 & 63.13 & 81.87 & 77.02 & 94.08 & 55.86 & 85.08 & 75.62\\
Swin-Unet \cite{cao2023swin}& 79.13& 21.55 & 85.47 & 66.53 & 83.28 & 79.61 & 94.29 & 56.58 & 90.66 & 76.60\\
nnFormer \cite{zhou2021nnformer}& 86.50 & 11.52 & 91.85 & 70.55 & \textbf{88.04} & 86.90 & 96.46 & 81.85 & 92.97 & 83.40\\
PHTrans \cite{liu2022phtrans}& 87.40 & 9.03 & 92.55 & 73.25 & 84.75 & \textbf{89.8} & 97.20 & \textbf{83.65} & 92.33 & 85.64\\
UNETR++ \cite{shaker2022unetr++}& 87.22 & 7.53 & 92.52 &71.25 & 87.54 & 87.18 & 96.42 & 81.10 & 95.77 & \textbf{86.01}\\
\hline
NexToU & \textbf{87.84} & \textbf{6.33} & \textbf{93.07} & \textbf{74.34} & 87.70 & 87.35 & 97.06 & 82.69 & \textbf{96.24} & 84.25\\
\hline
\end{tabular}
\end{table}
\subsection{Ablation Study}Table~\ref{tab4} presents an analysis of different components on the BTCV dataset. The "+PG" and "+SG" columns show that adding the Pool GNN and Swin GNN modules to the 3D U-Net model improved DSC by 1.48\% and 1.78\% and decreased HD by 1.80 mm and 3.52 mm, respectively. The results indicate that the BTI module can improve the DSC of NexToU with PG and SG by 0.61\% and reduce HD by 1.02 mm. Overall, incorporating the BTI module and the hybrid ViG-CNN architecture can significantly enhance the segmentation performance of medical images, demonstrating their effectiveness for this task.}
% tab 3 and tab4:
\begin{center}
\begin{minipage}{0.49\linewidth}
  \centering
  %\raggedright
  \captionof{table}{Results of segmentation on the RAVIR dataset.}\label{tab3}
  \begin{tabular}{c|c|ccc}
    \hline
    Methods & DSC$\uparrow$ & Artery & Vein\\
    \hline
    U-Net \cite{ronneberger2015u}& 72.67 & 70.36 & 74.97\\
    U-Net++ \cite{zhou2018unet++}& 75.35 & 73.21 & 77.48\\
    nnU-Net \cite{isensee2021nnu}& 77.54 & 74.85 & 80.23\\
    \hline
    NexToU & \textbf{78.21} & \textbf{75.62} & \textbf{80.79}\\
    \hline
    \end{tabular}
\end{minipage}\hfill
\begin{minipage}{0.49\linewidth}
  \centering
  %\raggedright
  \captionof{table}{Ablation analysis of different components on the BTCV dataset(params in M).}\label{tab4}
  \begin{tabular}{c|ccc}
    \hline
    Methods & Params & DSC$\uparrow$ & HD$\downarrow$\\
    \hline
    3D U-Net & 14.02 & 85.19 & 12.02 \\
    3D U-Net + PG & 18.54 & 86.67 & 10.22 \\
    3D U-Net + SG & 18.52 & 86.97 & 8.50 \\
    NexToU w/o BTI & 23.06 & 87.23 & 7.35 \\ 
    NexToU & 23.06 & \textbf{87.84} & \textbf{6.33}\\
    \hline
    \end{tabular}
\end{minipage}
\end{center}
\section{Conclusions} 
Our NexToU architecture for medical image segmentation, with innovative Pool GNN and Swin GNN modules, offers superior topological representation learning while minimizing computational costs. By leveraging the topological constraints among anatomical structures with BTI module, NexToU can efficiently and accurately differentiate structures based on their topological relationships. NexToU significantly improves the performance of medical image segmentation, as demonstrated by experimental results on various datasets. Furthermore, it offers a practical solution to achieve high-precision segmentation of complex anatomical structures, particularly the brain vasculature.
%\subsubsection{Acknowledgements} ***********************************************.


\newpage
\section{Supplementary Material} 
\begin{figure}[htb]
\centering
\includegraphics[width=0.67\textwidth]{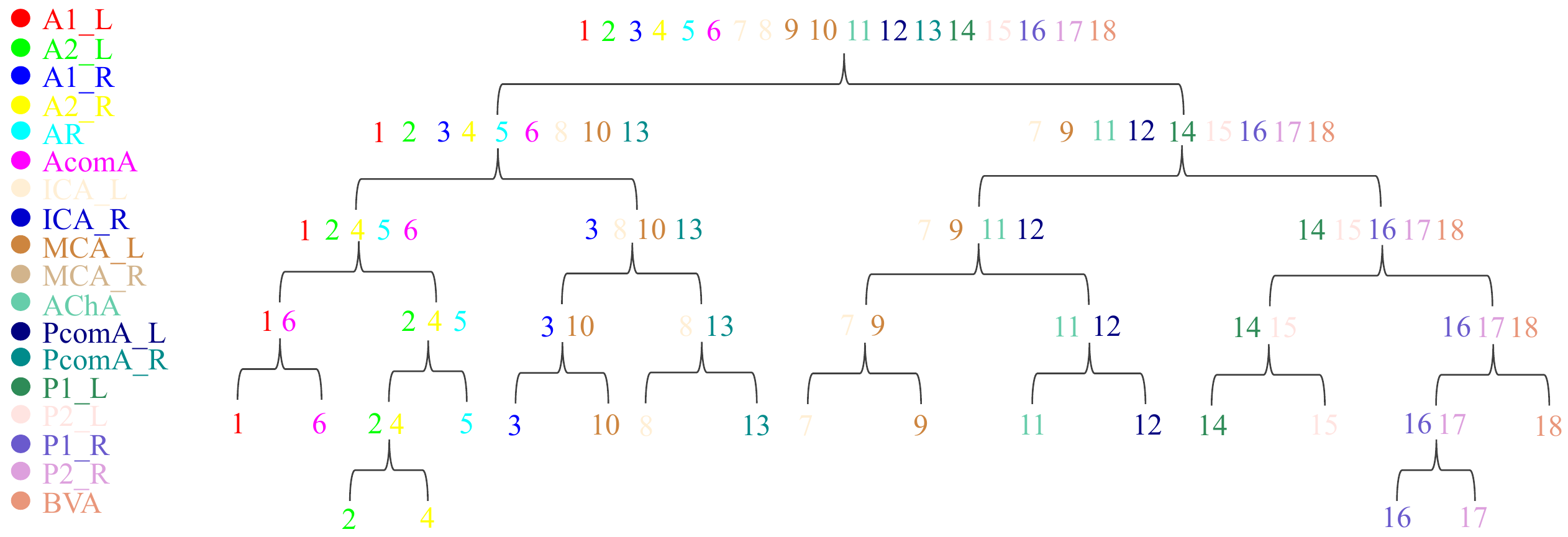}
\caption{The binary tree structure details of the ICA dataset in the BTI module.} \label{fig4}
\end{figure}
\begin{figure}[htb]
\centering
\includegraphics[width=0.67\textwidth]{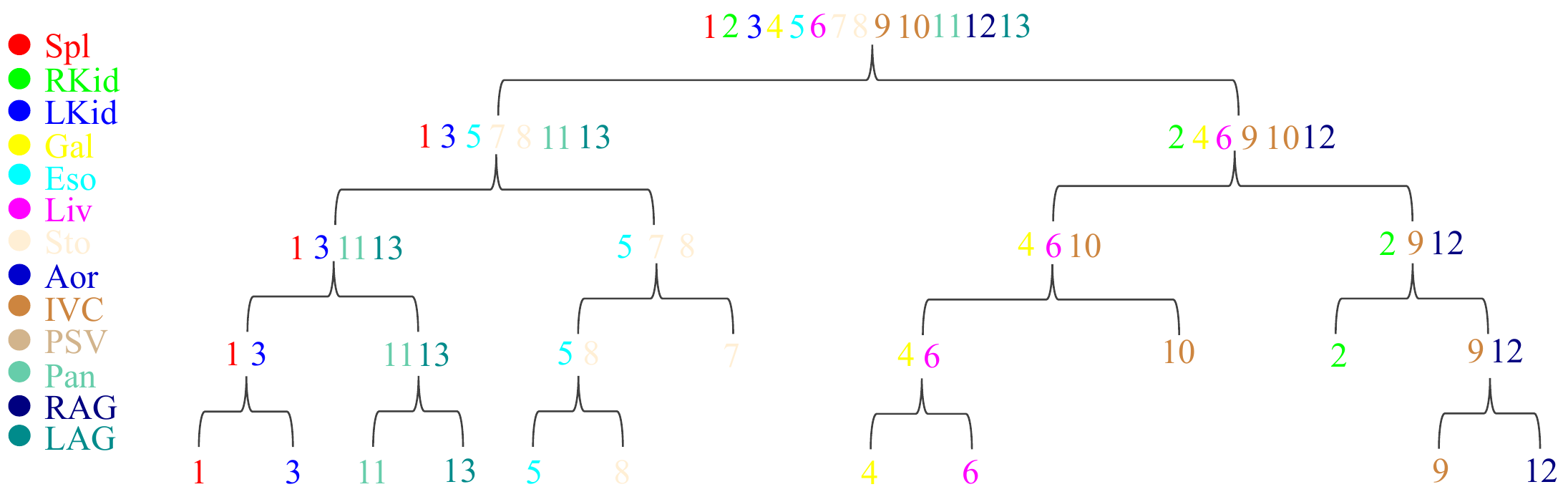}
\caption{The binary tree structure details of the BTCV dataset in the BTI module.} \label{fig5}
\end{figure}

%\subsection{Comparison of the TI Module and BTI Module Single-epoch Runtimes}
\begin{table}
\centering
\caption{Compare TI module and BTI module single-epoch runtimes on different datasets trained on a single NVIDIA-V100 GPU under identical conditions.}\label{tab5}
\begin{tabular}{c|c|c}
\hline
\multirow{2}{*}{Datasets} & \multirow{2}{*}{Methods} & \multirow{2}{*}{Runtimes (s)} \\
& & \\
\cline{1-3}
\multirow{2}{*}{ICA} & TI & 341.8 \\
& BTI & 217.7 \\
\hline
\multirow{2}{*}{BTCV} & TI & 203.4 \\
& BTI & 155.5 \\
\hline
\multirow{2}{*}{RAVIR} & TI & 79.2 \\
& BTI & 79.2 \\
\hline
\end{tabular}
\end{table}

%\subsection{Illustration of the Intracranial Arteries}
\begin{figure}[htb]
\centering
\includegraphics[width=0.32\textwidth]{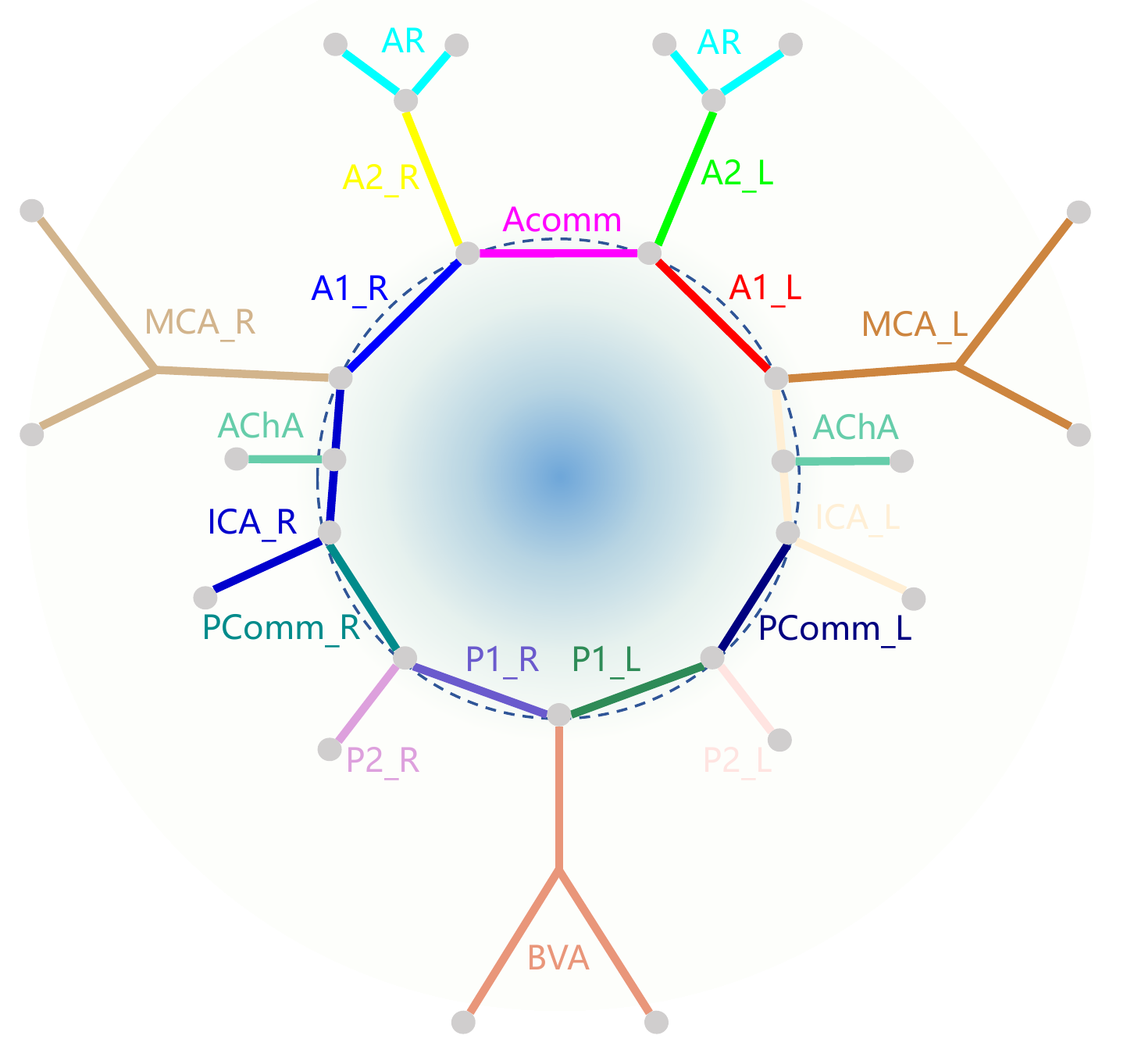}
\caption{The schematic illustration of the 18 diverse segments of intracranial arteries (ICA).} \label{fig6}
\end{figure}

%\subsection{The Full Name of Each Arterial Branch}
\begin{table}
\centering
\caption{The full name of each arterial branch in the ICA dataset.}\label{tab6}
\begin{tabular}{c|c}
\hline
Abbreviation & Full name\\
\hline
A1\_L & The first segment of the left anterior cerebral artery\\
A2\_L & The second segment of the left anterior cerebral artery\\
A1\_R & The first segment of the right anterior cerebral artery\\
A2\_R & The second segment of the right anterior cerebral artery\\
AR & The rest segment of the right anterior cerebral artery\\
AcomA & The anterior communicating artery\\
ICA\_L & The left internal carotid artery\\
ICA\_R & The right internal carotid artery\\
MCA\_L & The left middle cerebral artery\\
MCA\_R & The right middle cerebral artery\\
AchA & The anterior choroidal artery\\
PcomA\_L & The left posterior communicating artery\\
PcomA\_R & The right posterior communicating artery\\
P1\_L & The first segment of the left posterior cerebral artery\\
P2\_L & The second segment of the left posterior cerebral artery\\
P1\_R & The first segment of the right posterior cerebral artery\\
P2\_R & The second segment of the right posterior cerebral artery\\
BVA & The basilar and vertebral arteries\\
\hline
\end{tabular}
\end{table}

\end{document}